\documentclass[runningheads]{llncs}
\usepackage[T1]{fontenc}
\usepackage{graphicx}
\usepackage{amsmath,amssymb,amsfonts}
\usepackage{algorithmic}
\usepackage{textcomp}
\usepackage{xcolor}
\usepackage{multirow}
\usepackage{lipsum}
\usepackage{url}
\usepackage{tabularx}
\usepackage{booktabs}

\begin{document}
\title{Implementation and Performance Evaluation of TCP over QUIC Tunnels}
\author{Xuanhong Guo\orcidID{0009-0007-8772-6611} \and
Zekun Bao\orcidID{0009-0002-2755-2393} \and \\
Ying Chen*\orcidID{0000-0002-9520-3008}}
\authorrunning{X. Guo et al.}
\institute{Department of Computer Science, School of Artificial Intelligence\\Taizhou University, Taizhou 318000, Zhejiang, China\\
*Corresponding author: \email{ychen222@tzc.edu.cn}}
\maketitle
\begin{abstract}
QUIC, a UDP-based transport protocol, addresses several limitations of TCP by offering built-in encryption, stream multiplexing, and improved loss recovery. To extend these benefits to legacy TCP-based applications, this paper explores the implementation and evaluation of a TCP over QUIC tunneling approach. A lightweight, stream-based tunnel is constructed using the Rust-based Quinn library, enabling TCP traffic to traverse QUIC connections transparently. Performance is evaluated under varying network conditions, including packet loss, high latency, and out-of-order delivery. Results indicate that TCP over QUIC maintains significantly higher throughput than native TCP in lossy or unstable environments, with up to a high improvement under 20\% packet loss. However, under ideal network conditions, tunneling introduces modest overhead due to encryption and user-space processing. These findings provide insights into the trade-offs of TCP over QUIC tunneling and its suitability for deployment in dynamic or impaired networks.

\end{abstract}

\section{Introduction}
The evolution of transport protocols has been a driving force in the advancement of modern web applications. Traditional protocols like the Transmission Control Protocol (TCP), as defined in RFC 9293 \cite{rfc9293} and RFC 793 \cite{rfc793}, have long been the backbone of reliable data transmission over the Internet. However, with the increasing demand for low-latency, high-throughput, and secure communication, TCP’s limitations have become more pronounced \cite{9110324}. TCP’s reliance on a three-way handshake for connection establishment, susceptibility to head-of-line (HOL) blocking, and separate TLS implementation post-handshake introduce inefficiencies that hinder its performance in dynamic network environments.

To address these challenges, the QUIC (Quick UDP Internet Connections) protocol, initially developed by Google and standardized in RFC 9000 \cite{rfc9000}, has emerged as a promising alternative. QUIC leverages UDP as its transport layer, enabling a 0-RTT handshake for known servers, integrating TLS 1.3 \cite{rfc8446} for enhanced security \cite{ABDELHAFEZ2023101797}, and supporting stream multiplexing to mitigate HOL blocking. These features make QUIC particularly suitable for modern web applications such as HTTP/3, where low latency and high security are critical.

Despite QUIC’s advantages, its adoption in traditional TCP-based applications remains limited \cite{rfc9000}. Many legacy systems and applications rely heavily on TCP’s ordered delivery and reliability guarantees. This dichotomy raises the question: how can we harness QUIC’s benefits while maintaining compatibility with TCP-based applications? One potential solution is to encapsulate TCP traffic within QUIC tunnels, leveraging QUIC’s features while preserving the familiar semantics of TCP.

This paper explores the implementation and performance evaluation of a TCP over QUIC tunneling scheme. By encapsulating TCP traffic within QUIC packets, we aim to bridge the gap between traditional TCP applications and modern QUIC-based infrastructure. The study focuses on key performance metrics such as latency, throughput, and reliability across diverse network conditions. Through experimental evaluation, we demonstrate QUIC’s superiority in high-loss and fluctuating network environments, while also addressing challenges such as compatibility issues, computational overhead, and security concerns.

The remainder of this paper is structured as follows. Section \ref{sec:background} provides a detailed comparison of QUIC and TCP core mechanisms, along with an overview of network tunneling principles. Section \ref{sec:related_work} reviews related work on QUIC and TCP performance studies, as well as network tunneling research. Section \ref{sec:implementation} describes the implementation details of the TCP over QUIC tunneling scheme. Section \ref{sec:experimental_methodology} presents the experimental methodology. Section \ref{sec:results} presents evaluation metrics, then discusses the results of the performance evaluation. Finally, Section \ref{sec:conclusion} concludes the paper and outlines future research directions.

In this work, we implement a simple TCP over QUIC tunneling system and evaluate its performance under diverse network conditions. We emphasize that this work is exploratory in nature, focusing on empirical evaluation rather than proposing a new tunneling mechanism. Our contributions include a lightweight Rust baseline implementation, a thorough comparison with native TCP, and a detailed analysis of throughput, latency, and robustness. The results provide new insights into the feasibility and advantages of using QUIC as a transport layer for TCP-based applications.

\section{Background \& Motivation} \label{sec:background}

\subsection{Comparison of QUIC and TCP Core Mechanisms}

TCP, defined in RFC~9293 \cite{rfc9293} and RFC~793 \cite{rfc793}, has long underpinned reliable Internet transport. QUIC, standardized in RFC~9000 \cite{rfc9000}, addresses TCP’s latency and security limitations, particularly for web applications. Key distinctions include:

\begin{itemize}
    \item \textbf{Connection Establishment}: TCP requires a three-way handshake plus TLS 1.3 negotiation \cite{8538687}, incurring multiple RTTs. QUIC merges transport and TLS handshakes, enabling 0-RTT for known servers.
    \item \textbf{Multiplexing}: TCP’s single stream is prone to head-of-line blocking, while QUIC supports concurrent streams with independent flow control, reducing HOL delay \cite{kyratzis_quic_2022}.
    \item \textbf{Security}: TCP depends on separate TLS (RFC~8446), whereas QUIC integrates TLS~1.3, encrypting all packets and reducing latency \cite{fastly2021quic}.
    \item \textbf{Congestion Control}: Both employ algorithms such as New Reno (RFC~6582) \cite{rfc6582} and QUIC-specific variants (RFC~9002) \cite{rfc9002}, with QUIC’s user-space design allowing faster evolution.
    \item \textbf{Implementation}: TCP benefits from kernel optimizations, while QUIC’s user-mode operation adds per-packet overhead but increases flexibility \cite{fastly2021quic}.
\end{itemize}

\subsection{Network Tunneling Principle}

Network tunneling encapsulates one protocol in another for compatibility or security \cite{davie-stt-06}. For TCP over QUIC, two modes exist \cite{piraux-intarea-quic-tunnel-tcp-00}:

\begin{itemize}
    \item \textbf{Datagram Mode}: Encapsulates full packets (headers + payload) in QUIC datagrams \cite{piraux-intarea-quic-tunnel-tcp-00}, supporting arbitrary protocols but incurring high overhead.
    \item \textbf{Stream Mode}: Transmits TCP bytestreams over QUIC streams \cite{piraux-intarea-quic-tunnel-tcp-00}, reducing redundant headers via TLV negotiation (e.g., TCP Connect TLV) and aligning flow control with TCP windows. This mode is adopted in our implementation.
\end{itemize}

\section{Related Work} \label{sec:related_work}

\subsection{QUIC and TCP Performance Studies}

The performance of QUIC and TCP has been widely studied, especially for web and mobile applications requiring low latency and high throughput. QUIC, built atop UDP, integrates TLS~1.3, multiplexing, and 0-RTT setup, making it a strong TCP successor for latency-sensitive scenarios. Studies show QUIC incurs higher CPU overhead due to user-space processing and encryption, yet performs better than TCP in lossy or variable networks. 

For tunneling, encapsulating TCP over QUIC combines TCP’s compatibility with QUIC’s robustness. Recent work explores such hybrid systems to exploit QUIC’s loss recovery and multiplexing while preserving TCP semantics. Our study extends this line by implementing and experimentally evaluating a TCP over QUIC tunnel in Rust (Quinn), validating performance trade-offs under controlled loss and latency.

\subsection{Comparison of QUIC Implementations}

QUIC implementations vary in language, features, and performance. Hertelli et al.~\cite{hertelli_comparison_2022} compare Chromium-quic “Quiche”, Cloudflare “Quiche”, Mvfst, MsQuic, Quic-go, Aioquic, and Haskell-quic, noting differences in maintenance, congestion control (e.g., CUBIC, NewReno, HyStart++), and adherence to IETF standards.

Marti~\cite{marti_continuous_2024} benchmarked MsQuic and Quinn. Optimizing Quinn with larger segment sizes nearly doubled throughput and reduced system calls, but MsQuic consistently outperformed Quinn under loss or reordering. These results highlight Quinn’s efficiency under stable networks and MsQuic’s robustness under instability. Given Quinn’s strengths and its Rust ecosystem integration, we adopt it for our TCP over QUIC tunnel.

\subsection{Identified Deficiencies}

Although there have been some related studies such as rstun \cite{rstun}, their design mainly emphasizes feature richness and product usability. However, for the purposes of controlled experiments, this relative complexity introduces additional variables that may obscure the essential performance characteristics of TCP over QUIC tunnels. Therefore, we developed a minimal but fully functional Rust-based baseline approach stripped of unnecessary mechanisms to ensure clarity and repeatability.

On top of this baseline, we conducted systematic evaluations across diverse network conditions. Unlike prior work, which primarily compared QUIC and TCP directly, our study isolates the tunneling layer itself, allowing us to quantify trade-offs such as CPU overhead, throughput under loss, and robustness in high-latency environments.

\section{System Design \& Implementation} \label{sec:implementation}

\begin{figure*}[ht]
    \centering
    \includegraphics[width=\linewidth]{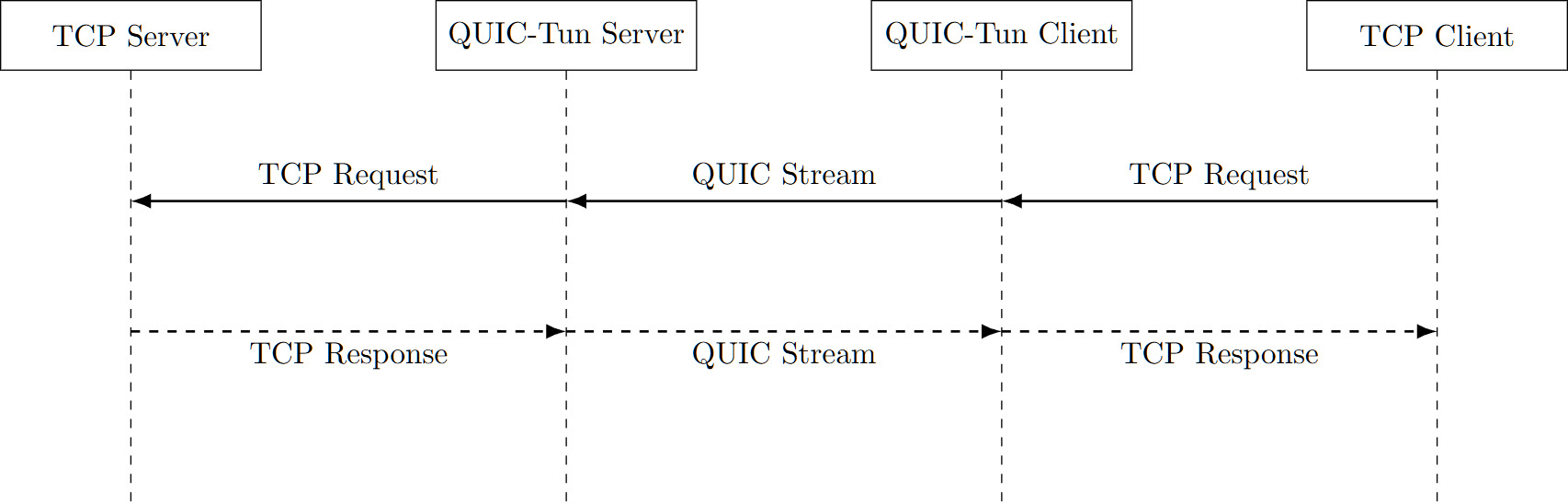}
    \caption{QUIC-Tun architecture: traffic flow from a local TCP client through the QUIC tunnel to a remote TCP server.}
    \label{fig:quic-tun_squence_diagram}
\end{figure*}

The TCP over QUIC tunneling system consists of three core components: a bidirectional data transfer utility, a server-side tunnel endpoint, and a client-side TCP–QUIC proxy. The implementation is in Rust using the Tokio runtime \cite{tokio} and the Quinn QUIC library \cite{quinn}.

\subsection{Bidirectional Transfer}

The central \texttt{bidirectional\_copy} utility establishes full-duplex forwarding between TCP and QUIC streams. It runs two concurrent tasks to relay traffic in both directions, synchronizes their lifecycle, and ensures clean shutdown. Rust’s ownership model and async runtime guarantee atomicity and safe resource management.

\subsection{Server-Side Endpoint}

The QUIC server operates as a persistent endpoint. Using Quinn’s \texttt{Endpoint}, it accepts connections and spawns isolated async tasks. Each QUIC stream is mapped to a TCP connection via the transfer utility. Configuration includes self-signed certificates (via \texttt{rcgen}), a maximum of 100 concurrent bidirectional streams, disabled unidirectional streams, and a 2s keep-alive interval. TLS compatibility is preserved.

\subsection{Client-Side Proxy}

The client listens on a local TCP port and establishes a QUIC connection upon demand. Each TCP stream is mapped to a bidirectional QUIC stream. Security options include standard TLS verification or a relaxed mode (for testing). Keep-alive parameters match the server. Each session runs in an async task with error logging.

\subsection{Integration}

A CLI interface (via Clap \cite{clap}) provides unified commands:
\begin{itemize}
    \item Server: \texttt{quic-tun serve --dest <TCP\_ADDR> --bind <QUIC\_ADDR>}
    \item Client: \texttt{quic-tun connect --dest <QUIC\_ADDR> --bind <TCP\_ADDR>}
\end{itemize}

Cryptographic operations rely on Rustls \cite{rustls} with the ring provider \cite{ring}. The modular design supports extensibility to other tunneling scenarios while maintaining throughput and reliability.

\section{Experimental Methodology} \label{sec:experimental_methodology}

This section outlines the methodology used to evaluate TCP over QUIC tunneling.

\subsection{Experimental Setup}

Experiments were conducted in a controlled virtual environment for repeatability. The setup included:

\begin{itemize}
    \item \textbf{Virtual Machines}: Two Debian 12.10.0 VMs on the same host, each with 4 vCPUs (host-passthrough) and 8 GB RAM.
    \item \textbf{OS/Kernel}: Debian GNU/Linux 12.10.0 with Linux 6.1.
    \item \textbf{Protocol Implementations}:
    \begin{itemize}
        \item TCP: Default Linux TCP stack using Cubic congestion control
        \item QUIC: Implemented via the Quinn library (Rust)
    \end{itemize}
    \item \textbf{Performance Tool}: \texttt{iperf3}, repeated 5 times per protocol with default parameters, measuring throughput (Mbps) and CPU usage.
    \item \textbf{Monitoring}: \texttt{tcpdump} for traffic capture and validation.
\end{itemize}

Units: Bandwidth in MB/s, throughput in Mbps (\(1~\text{MB/s} = 8~\text{Mbps}\)).

\subsection{Evaluation Metrics}

\begin{itemize}
    \item \textbf{Throughput}: $D / t$, where $D$ is data (bits), $t$ is transfer time (s).
    \item \textbf{CPU Usage}: $(C \times P) / T \times 100\%$, where $C$ = CPU cycles per packet, $P$ = packets/sec, $T$ = total CPU cycles/sec.
\end{itemize}

Measuring RTT in this context is unnecessary because tunnel effects appear in throughput, response time, or CPU usage. Moreover, TCP-layer RTT within a QUIC tunnel is distorted by QUIC’s congestion control and retransmission, making it unsuitable as a standalone metric.

\section{Results \& Analysis} \label{sec:results}

\subsection{Bandwidth Sensitivity}

\begin{table*}
    \centering
    \caption{Comparison of performance metrics under varying bandwidth}
    \label{tab:bandwidth_metrics}
    \begin{tabularx}{\textwidth}{|c|*{8}{>{\centering\arraybackslash}X|}}
\hline
\multirow{3}{*}{\textbf{Bandwidth (MB/s)}} 
    & \multicolumn{4}{c|}{\textbf{QUIC Tunnel}} 
    & \multicolumn{4}{c|}{\textbf{Plain TCP}} \\
\cline{2-9}
    & \multicolumn{2}{c|}{\textbf{Bitrate (Mbps)}} 
    & \multicolumn{2}{c|}{\textbf{CPU (\%)}} 
    & \multicolumn{2}{c|}{\textbf{Bitrate (Mbps)}} 
    & \multicolumn{2}{c|}{\textbf{CPU (\%)}} \\
\cline{2-9}
    & Sndr & Rcvr
    & Sndr & Rcvr 
    & Sndr & Rcvr 
    & Sndr & Rcvr \\
\hline
100 & 132   & 126   & 0.3 & 17.6 & 824  & 820  & 6.5 & 15.3 \\
\hline
10  & 76.0  & 70.1  & 0.5 & 5.8  & 76.5 & 73.2 & 0.6 & 1.2  \\
\hline
1   & 11.8  & 6.18  & 0.3 & 1.1  & 11.1 & 7.75 & 0.3 & 0.2  \\
\hline
\end{tabularx}
\end{table*}

Table~\ref{tab:bandwidth_metrics} compares throughput under varying bandwidth. As expected, native TCP outperforms the tunnel under ideal conditions, benefiting from kernel-level optimizations. Since the tunnel is designed for degraded networks, subsequent experiments adopt a default bandwidth of 10~MB/s.

\subsection{Packet Loss Resilience}

We evaluated both approaches at different packet loss rates using iperf3. The packet loss rate is defined as:
\[
\text{Packet Loss Rate} = \frac{N_{\text{lost}}}{N_{\text{sent}}} \times 100\%.
\]

\begin{table*}
    \centering
    \caption{Comparison of performance metrics under varying packet loss rates}
    \label{tab:packet_loss_metrics}
    \begin{tabularx}{\textwidth}{|c|*{8}{>{\centering\arraybackslash}X|}}
\hline
\multirow{3}{*}{\textbf{Loss (\%)}} 
    & \multicolumn{4}{c|}{\textbf{QUIC Tunnel}} 
    & \multicolumn{4}{c|}{\textbf{Plain TCP}} \\
\cline{2-9}
    & \multicolumn{2}{c|}{\textbf{Bitrate (Mbps)}} 
    & \multicolumn{2}{c|}{\textbf{CPU (\%)}} 
    & \multicolumn{2}{c|}{\textbf{Bitrate (Mbps)}} 
    & \multicolumn{2}{c|}{\textbf{CPU (\%)}} \\
\cline{2-9}
    & Sndr & Rcvr
    & Sndr & Rcvr 
    & Sndr & Rcvr 
    & Sndr & Rcvr \\
\hline
0   & 72.7 & 68.0 & 0.5 & 9.0 & 84.9 & 81.5 & 0.9 & 1.6 \\
\hline
5   & 58.3 & 53.1 & 0.1 & 11.1 & 47.5 & 46.6 & 0.3 & 0.8 \\
\hline
10  & 32.2 & 26.9 & 0.6 & 5.8 & 19.5 & 19.1 & 0.3 & 0.2 \\
\hline
15  & 18.1 & 13.2 & 0.1 & 2.3 & 5.14 & 4.93 & 0.1 & 0.0 \\
\hline
20  & 11.1 & 5.97 & 0.2 & 0.9 & 2.41 & 2.18 & 0.3 & 0.1 \\
\hline
\end{tabularx}
\end{table*}

\begin{figure}[ht]
    \centering
    \includegraphics[width=0.8\textwidth]{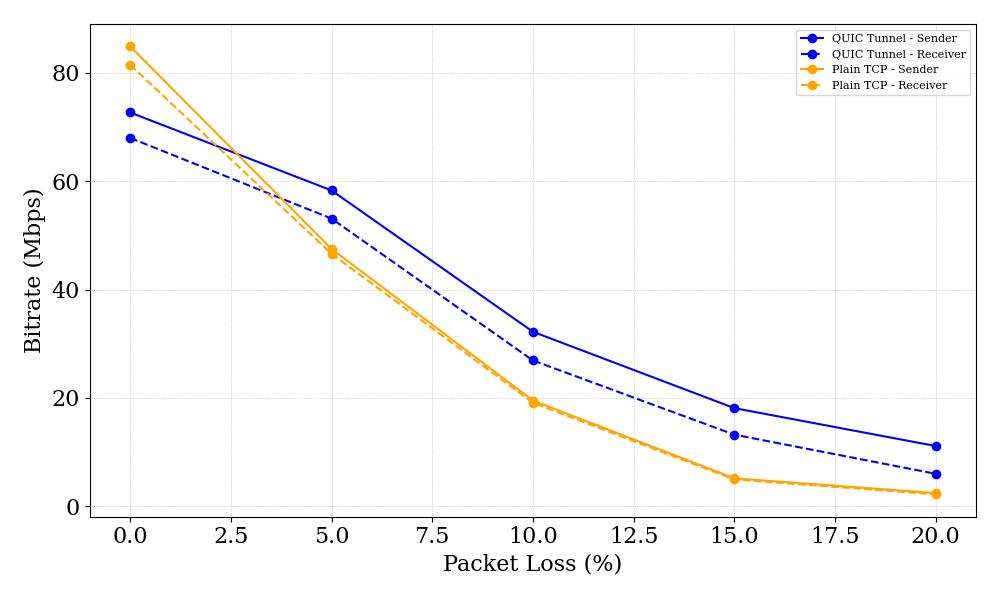}
    \caption{This figure shows the impact of the two connections on throughput at different packet loss rates}
    \label{bitrate_vs_packet_loss}
\end{figure}

Table~\ref{tab:packet_loss_metrics} and Fig.~\ref{bitrate_vs_packet_loss} show that:
\begin{itemize}
    \item TCP is slightly better at low loss ($\leq$5\%) due to tunnel overhead.
    \item Beyond 5\%, the tunnel consistently outperforms TCP, e.g., 4.6$\times$ higher throughput at 20\% loss.
    \item QUIC degrades more gracefully, highlighting its efficient recovery mechanisms.
\end{itemize}

The tunnel incurs somewhat higher CPU usage, mainly from encryption, user–kernel transitions, and encapsulation.

\subsection{Latency Impact}

Latency is defined as:
\[
\text{Latency} = \sum_{i=1}^{N} (T_{\text{prop}}^i + T_{\text{trans}}^i + T_{\text{queue}}^i + T_{\text{proc}}^i),
\]
with an extra $T_{\text{QUIC}}$ term for tunnel overhead. 

\begin{table*}
    \centering
    \caption{Comparison of performance metrics under varying lags}
    \label{tab:lag_metrics}
    \begin{tabularx}{\textwidth}{|c|*{8}{>{\centering\arraybackslash}X|}}
\hline
\multirow{3}{*}{\textbf{Lag (ms)}}
    & \multicolumn{4}{c|}{\textbf{QUIC Tunnel}} 
    & \multicolumn{4}{c|}{\textbf{Plain TCP}} \\
\cline{2-9}
    & \multicolumn{2}{c|}{\textbf{Bitrate (Mbps)}} 
    & \multicolumn{2}{c|}{\textbf{CPU (\%)}} 
    & \multicolumn{2}{c|}{\textbf{Bitrate (Mbps)}} 
    & \multicolumn{2}{c|}{\textbf{CPU (\%)}} \\
\cline{2-9}
    & Sndr & Rcvr
    & Sndr & Rcvr 
    & Sndr & Rcvr 
    & Sndr & Rcvr \\
\hline
0    & 77.6 & 70.3 & 0.8 & 8.1 & 66.1 & 53.3 & 0.6 & 0.6 \\
\hline
50   & 59.7 & 54.7 & 0.3 & 6.8 & 61.8 & 56.1 & 1.2 & 0.8 \\
\hline
100  & 59.4 & 53.8 & 0.3 & 5.2 & 50.7 & 37.8 & 0.6 & 0.9 \\
\hline
200  & 37.0 & 31.5 & 0.3 & 3.4 & 43.2 & 37.2 & 0.7 & 0.6 \\
\hline
500  & 15.5 & 9.97 & 0.1 & 1.5 & 4.72 & 3.23 & 0.0 & 0.0 \\
\hline
1000 & 6.60 & 2.76 & 0.1 & 0.0 & 0.73 & 0.61 & 0.1 & 0.0 \\
\hline
\end{tabularx}
\end{table*}

\begin{figure}[ht]
    \centering
    \includegraphics[width=0.8\textwidth]{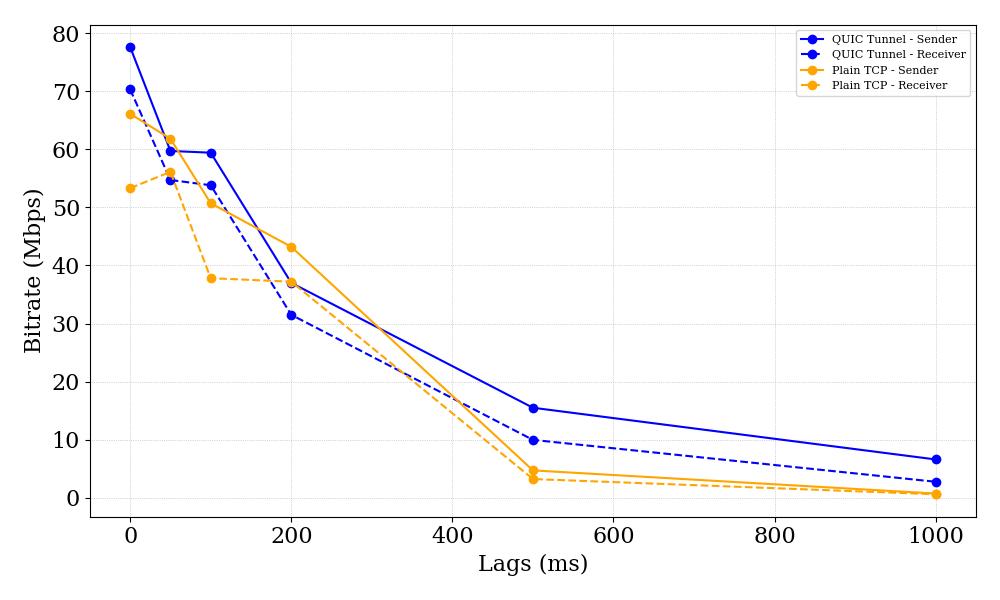}
    \caption{This figure shows the impact of the two connections on throughput at different lags}
    \label{bitrate_vs_lag}
\end{figure}

From Table~\ref{tab:lag_metrics} and Fig.~\ref{bitrate_vs_lag}:
\begin{itemize}
    \item Throughput decreases with latency for both protocols, but the tunnel maintains higher throughput at moderate latencies.
    \item At 1000~ms, the tunnel is still clearly ahead (6.60 vs. 0.73~Mbps on sender side).
    \item CPU usage varies, with the tunnel often slightly higher, but not consistently better or worse.
\end{itemize}

\subsection{Packet Reordering Tolerance}

Out-of-order rate is defined as:
\[
\text{Out-of-Order Rate} = \frac{N_{\text{out-of-order}}}{N_{\text{total}}} \times 100\%.
\]

\begin{table*}
    \centering
    \caption{Comparison of performance metrics under varying out of order rates}
    \label{tab:out_of_order_metrics}
    \begin{tabularx}{\textwidth}{|c|*{8}{>{\centering\arraybackslash}X|}}
\hline
\multirow{3}{*}{\textbf{Out of Order (\%)}} 
    & \multicolumn{4}{c|}{\textbf{QUIC Tunnel}} 
    & \multicolumn{4}{c|}{\textbf{Plain TCP}} \\
\cline{2-9}
    & \multicolumn{2}{c|}{\textbf{Bitrate (Mbps)}} 
    & \multicolumn{2}{c|}{\textbf{CPU (\%)}} 
    & \multicolumn{2}{c|}{\textbf{Bitrate (Mbps)}} 
    & \multicolumn{2}{c|}{\textbf{CPU (\%)}} \\
\cline{2-9}
    & Sndr & Rcvr
    & Sndr & Rcvr 
    & Sndr & Rcvr 
    & Sndr & Rcvr \\
\hline
0   & 74.8 & 69.9 & 0.5 & 8.8 & 44.0 & 40.6 & 0.6 & 0.8 \\
\hline
5   & 76.4 & 71.5 & 0.6 & 8.7 & 44.9 & 40.8 & 0.3 & 0.9 \\
\hline
10  & 77.9 & 70.1 & 0.6 & 9.2 & 44.1 & 40.8 & 0.9 & 0.8 \\
\hline
15  & 80.1 & 75.1 & 0.5 & 9.1 & 52.2 & 48.8 & 0.8 & 0.6 \\
\hline
20  & 76.2 & 69.3 & 1.0 & 7.0 & 52.3 & 48.9 & 0.5 & 1.4 \\
\hline
\end{tabularx}
\end{table*}

\begin{figure}[ht]
    \centering
    \includegraphics[width=0.8\textwidth]{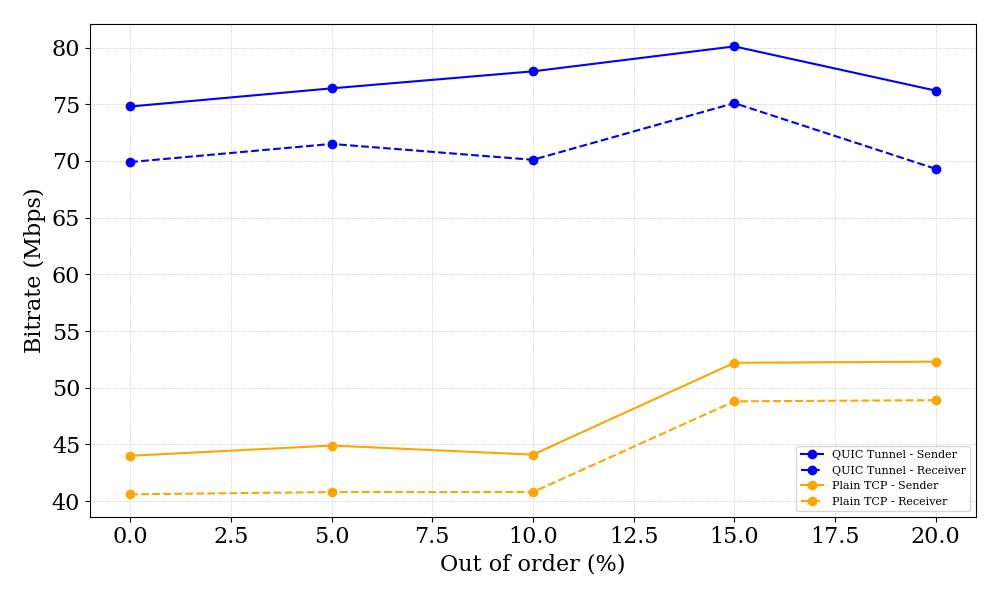}
    \caption{This figure shows the impact of the two connections on throughput at different out of order rates}
    \label{bitrate_vs_out_of_order}
\end{figure}

As shown in Table~\ref{tab:out_of_order_metrics} and Fig.~\ref{bitrate_vs_out_of_order}:
\begin{itemize}
    \item QUIC tolerates reordering well due to its stream-based design.
    \item TCP over QUIC inherits some benefit but remains constrained by TCP’s in-order semantics.
    \item The tunnel shows slightly better stability than native TCP under reordering.
\end{itemize}

Finally, note that both inner TCP and outer QUIC implement congestion control independently. Their interaction may cause inefficiencies such as oscillation, which require further investigation beyond our VM-based tests.

\section{Conclusion \& Future Work} \label{sec:conclusion}

This paper presents the implementation and evaluation of a TCP-based tunneling scheme over QUIC, with the aim of leveraging the advantages of QUIC while maintaining compatibility with traditional TCP-based applications. The results of comprehensive experiments demonstrate that under lossy or high-latency network conditions, TCP over QUIC offers significant resilience and performance advantages compared to native TCP connections. These benefits are primarily attributed to the inherent features of QUIC, including stream multiplexing and efficient loss recovery mechanisms.

However, our evaluation also reveals the limitations of TCP over QUIC in high-quality network environments. In such conditions, the dual-protocol stack introduces additional computational overhead and latency due to user-space processing, encryption, and protocol encapsulation. This overhead leads to lower throughput and higher CPU usage compared to direct TCP, which benefits from kernel-level optimizations. As such, while TCP over QUIC is highly effective in degraded or mobile network scenarios, it is less efficient in well-provisioned, stable networks.

Our current evaluation is limited to a single QUIC implementation (Quinn) and VM-based setups. Broader validation across diverse QUIC stacks and real-world WAN environments will further strengthen the conclusions.

Further research could explore more forms of QUIC-based network tunneling, such as TCP over WebTransport \cite{ietf-webtrans-http3-12}, to further develop the potential of QUIC in the transformation of the network protocol. In addition, TCP over QUIC tunnel can optimize the processing of different application layer protocols and improve the performance of traditional application layer protocols under this tunnel. Future work should also investigate the congestion control interactions between the inner TCP and the outer QUIC layers, as this remains a fundamental challenge for tunneling designs.

\section{Data Availability}

The data sets generated and analyzed during the current study are not persistently stored, as all experimental data were dynamically generated during runtime. Due to variations in network conditions, system load, and inherent randomness, storing raw data such as packet captures or instantaneous performance metrics was deemed unnecessary. To ensure reproducibility, all testing methods and the full source code used in this study are publicly available at the GitHub repository: \url{https://github.com/ElaBosak233/quic-tun}. Researchers can reproduce the results by building and executing the code under similar or customized network conditions.

\bibliographystyle{splncs04}
\bibliography{ref}
\end{document}